\documentclass[useAMS,usenatbib,fleqn]{mn2e}

\usepackage{times}
\usepackage{amsmath}
\usepackage{mathrsfs}
\usepackage{amssymb}
\usepackage{amsbsy}
\usepackage{graphicx}
\usepackage{subfigure}
\usepackage{paralist}
\usepackage{color}

\usepackage{aas_macros}
\usepackage{hyperref}
\graphicspath{{plots/}}

\newcommand{\teff}{T$_{\rm eff}$}
\newcommand{\logg}{$\log g$}
\newcommand{\meta}{\hbox{[M/H]}}

\def\kms{\,{\rm km~s^{-1}}}
\def\d{{\rm d}}\def\kpc{\,{\rm kpc}}
\def\rg{r_{\rm g}}\def\rp{r_{\rm p}}\def\ra{r_{\rm a}}
\def\rb{\overline{r}}\def\rb{\overline{r}}
\def\vc{v_{\rm c}}
\def\dex{\,{\rm dex}}
\def\Myr{\,{\rm Myr}}\def\Gyr{\,{\rm Gyr}}
\title[Evidence from RAVE for Stellar Radial Migration]
  {The Rich Are Different: \\Evidence from the RAVE Survey for Stellar Radial Migration}
\author[G.~Kordopatis et al.]
  {	G.~Kordopatis,$^{1,2}\thanks{E-mail:~gkordo@ast.cam.ac.uk}$
	J.~Binney,$^3$
  	G.~Gilmore,$^1$
	R.F.G.~Wyse,$^4$
	V.~Belokurov,$^1$	
	P.J.~McMillan,$^3$	
	\newauthor{
	P.~Hatfield,$^{5,1}$	
	E.~K.~Grebel,$^6$
	M.~Steinmetz,$^2$
	J.F.~Navarro,$^{7}$
	G.~Seabroke,$^{8}$
	I.~Minchev,$^2$
	}
			\newauthor{
			C.~Chiappini,$^2$
			O.~Bienaym\'e,$^9$
			J.~Bland-Hawthorn,$^{10}$
			K.C.~Freeman,$^{11}$
			B.~K.~Gibson,$^{12,13}$
			}
			\newauthor{
			A.~Helmi,$^{14}$
			U.~Munari,$^{15}$
			Q.~Parker,$^{16, 17,18}$			
			W.A.~Reid,$^{16,17}$
			A.~Siebert,$^{8}$ 
			A.~Siviero,$^{19}$
			T. Zwitter$^{20}$
			}\\ 
  $^1$~Institute of Astronomy, University of Cambridge, Madingley Road, Cambridge CB3 0HA, United Kingdom\\
   $^2$ Leibniz-Institut f\"ur  Astrophysik Potsdam (AIP), An der Sternwarte 16, 14482 Potsdam, Germany\\
    $^3$~Rudolf Peierls Centre for Theoretical Physics, Keble Road, Oxford, OX1 3NP, United Kingdom \\
  $^4$~Johns Hopkins University, Homewood Campus, 3400 N Charles Street, Baltimore, MD 21218, USA\\
  $^5$~Department of Physics, University of Oxford, Denys Wilkinson Building, Keble Road, Oxford, OX1 3RH, United Kingdom \\
   $^{6}$ {Astronomisches Rechen-Institut, Zentrum f\"ur Astronomie der Universit\"at Heidelberg, M\"onchhofstr.\ 12--14, 69120 Heidelberg, Germany}\\
  $^{7}$~CIfAR Senior Fellow, Department of Physics and Astronomy, University of Victoria, Victoria BC, Canada V8P 5C2\\
 $^{8}$ Mullard Space Science Laboratory, University College London, Holmbury St Mary, Dorking, RH5 6NT, United Kingdom\\
   $^{9}$~Observatoire astronomique de Strasbourg, Universit\'e de Strasbourg, CNRS, UMR 7550, 11 rue de l'Universit\'e, F-67000 Strasbourg, France\\
  $^{10}$~Sydney Institute for Astronomy, School of Physics A28, University of Sydney, NSW 2611, Australia \\
  $^{11}$~RSAA Australian National University, Mount Stromlo Observatory, Cotter Road, Weston Creek, Canberra, ACT 72611, Australia\\
  $^{12}$~Institute for Computational Astrophysics, Dept of Astronomy \& Physics, Saint MaryÕs University, Halifax, NS, BH3 3C3, Canada\\
  $^{13}$~Jeremiah Horrocks Institute, University of Central Lancashire, Preston, PR1 2HE, United Kingdom\\
  $^{14}$~Kapteyn Astronomical Institute, University of Groningen, P.O. Box 800, 9700 AV Groningen, The Netherlands\\
  $^{15}$~INAF Astronomical Observatory of Padova, 36012 Asiago (VI), Italy \\
  $^{16}$ Department of Physics and Astronomy, Macquarie University, Sydney, NSW 2109 Australia\\
$^{17}$ Research Centre in Astronomy, Astrophysics and Astrophotonics, Macquarie University, Sydney, NSW 2109 Australia \\
$^{18}$ Australian Astronomical Observatory, PO Box 915, North Ryde NSW 1670, Australia\\
$^{19}$~Department of Physics and Astronomy, Padova University, Vicolo dellÕOsservatorio 2, I-35122 Padova, Italy\\
$^{20}$ Faculty of Mathematics and Physics, University of Ljubljana, 1000 Ljubljana, Slovenia
}

\pagerange{\pageref{firstpage}--\pageref{lastpage}} \pubyear{2014}

\def\LaTeX{L\kern-.36em\raise.3ex\hbox{a}\kern-.15em
    T\kern-.1667em\lower.7ex\hbox{E}\kern-.125emX}

\begin{document}

\label{firstpage}

\maketitle

\begin{abstract}
Using the \emph{RAdial Velocity Experiment} fourth data release (RAVE DR4),
and a new metallicity calibration that will be also taken into account in the
future RAVE DR5, we investigate the existence and the properties of
super-solar metallicity stars ($\meta \gtrsim+0.1\dex$) in the sample, and
in particular in the Solar neighbourhood. We find that RAVE is rich
in super-solar metallicity stars, and that the local metallicity
distribution function declines remarkably slowly up to $+0.4\dex$.
 Our results show that the kinematics and height distributions of the
super-solar metallicity stars are identical to those of the $\meta\lesssim0$
thin-disc giants that we presume were locally manufactured.  The
eccentricities of the super-solar metallicity stars indicate that half of
them are on a roughly circular orbit ($e \leq 0.15$), so under the
 assumption that the metallicity of the interstellar medium at a given radius never
decreases with time, they must have increased their angular momenta by
scattering at corotation resonances of spiral arms from regions far
inside the Solar annulus. The likelihood
that a star will migrate radially does not seem to decrease significantly
with increasing amplitude of vertical oscillations within range of
oscillation amplitudes encountered in the disc.

\end{abstract}


%%%%%%%%%%%%%%%%%%%%%%%%%%%%
\section{Introduction}

The disc is our Galaxy's dominant visible component and contains
most of the baryons that lie within a sphere of radius $r\sim100\kpc$. Hence if we are
to understand how our very typical Galaxy has arisen within the $\Lambda$CDM
paradigm \citep{Springel06}, we need to know how the disc is structured,
functions and was formed. Since the seminal works of
\cite{Spitzer53,Pagel75,Matteucci89} and others, the relevant framework
has been recognised to be the accretion of cool gas onto a centrifugally
supported disc within which stars form on nearly circular orbits. Dying stars
enrich the star-forming gas with metals, and fluctuations in the Galaxy's
gravitational field cause stars to migrate to less circular orbits that are
more inclined to the Galactic plane \citep[][and references
therein]{Binney13,Sellwood14}.

In a classic study \citet{Eggen62} showed that the orbits of the stars and
the chemical composition of their atmospheres suggest how one may reconstruct
the history of the Milky Way. While the chemical composition of a star
retains (to a good approximation) an imprint of the chemistry of the
Inter-Stellar Medium (ISM) at the time and place of its birth
\citep[e.g.:][]{Yoshii81, Freeman02}, its orbit depends both on the
environment in which the star formed (within circularly orbiting gas or gas
in free fall, or the gas disc of a satellite) and subsequent evolution of the
orbit in response to fluctuations in the Galaxy's gravitational field
(generated by molecular clouds, spiral structure, the bar, halo substructure,
etc). The correlations between chemistry and kinematics are therefore tracers
of the coevolution of nucleosynthesis and dynamical evolution.

The component $L_z$ of angular momentum about the Galaxy's approximate
symmetry axis plays a crucial role, and with $L_z$ we associate a
guiding-centre radius $\rg$ by the equation
$L_z=\rg\vc(\rg)$, where $\vc(r)$ is the speed of a circular orbit at $r$. A
star with angular momentum $L_z$ executes radial oscillations around $\rg$.
Changes in $L_z$ are therefore associated with changes in $\rg$ and one
speaks of ``radial migration'' when $L_z$ changes.

Measurements of the abundances and metallicities\footnote{We denote the
overall metallicity of a star, $\meta$,  the ratio of the abundance of any element to its abundance in the Sun. It is defined as $\meta =\log({\rm M/H})_\star - \log({\rm M/H})_\odot$.} of young
O and B stars, and nebular abundances, reveal the chemistry of the current
ISM, whereas measurements of lower-mass FGK stars, reveal the chemistry of
the ISM billions of years in the past.  The available data are consistent
with the conjecture that within a given Galactocentric annulus the ISM is
chemically homogeneous. In particular, the ISM is very homogeneous within
several hundred parsecs of the Sun \citep{Cartledge06}. So if we could
establish the ISM's radial metallicity profile $\meta_\tau(R)$ for each time
$\tau$ in the past, we could infer the birth radius of a star of age $\tau$
from its measured value of $\meta$ \citep[see][and references
therein for radial metallicity gradients of FGK stars]{Gazzano13,Boeche13b,Hayden14}. 

Stars in the solar neighbourhood with metallicities
above solar ($\meta>0$) are especially powerful probes of the evolution of
the Milky Way's disc. They can form only after several previous generations
of stars have enriched their local ISM \citep[e.g.][]{Pagel97,Matteucci03}.  This can either take several billion
years in regions where the star formation rate is low and approximately
constant with time \citep[OB stars in the Milky Way disc near the Sun only
reach, on average, $\meta=0$,][]{Nieva12}, or occur rapidly in dense
environments, such as the Galactic bulge, where we find stars with $\meta>0$ that are
several billion years old \citep[e.g.:][]{Whitford83,Hill11}. 

Given the homogeneity of the ISM, stars with $\meta\gtrsim0.15\dex$ (noted
super metal-rich stars, SMR, hereafter) must have formed inside the Sun's
Galactocentric distance $R_0$, and we see them here either because they have
significantly increased their Galactocentric angular momenta, and thus their
guiding-centre radii \citep{Grenon89, Grenon99,Chiappini09}, or because they have moved
to significantly eccentric orbits, or on account of a combination of both
these processes. 

\cite{Sellwood02} showed that the dominant effect of transient spiral
structure is to cause stars that are in corotation resonance (CR) with the
spiral to exchange angular momentum without changing the eccentricity of
their orbits.  They dubbed this process ``churning''. \cite{Lynden-Bell72} had
already shown that stars that are in a Lindblad resonance exchange angular
momentum with the spiral in such a way that on average they move to more
eccentric orbits: at inner Lindblad resonance (ILR) stars typically surrender
angular momentum to the spiral, while at outer Lindblad resonance (OLR) they
gain angular momentum from the spiral. Since stars scattered at ILR lose
angular momentum, and the great majority of disc stars were born inside $R_0$
(which is of order 3 of the disc's exponential scale lengths), SMR stars on
highly eccentric orbits are likely to have been scattered at OLR and thus
have increased both their angular momenta and eccentricity.

Hence a star that has markedly increased its angular momentum without moving
to a highly eccentric orbit must have been scattered at CR by the churning
process.  The azimuthal velocities of these stars will not lag the circular
velocity by much. On the other hand, stars that reach the Sun on eccentric orbits from
guiding-centres that are significantly smaller than $R_0$ will lag the local
circular speed significantly.  Hence by measuring the random velocities and
the azimuthal speeds of stars we should be able to determine the relative
importance of scattering at CR and at OLR.

The observational evidence for radial migration is still relatively scanty.  In nearby disc galaxies, \citet{Yoachim12} and \citet{Radburn-Smith12} 
measured for a subsample of their targets a change of the age of the dominant population at the location of the break in the disc surface brightness, in agreement with the theoretical work of \citet{Roskar08} on broken exponential profile in galaxies experiencing radial migration.  
As far as the Milky Way is concerned,  
\citet{Sellwood02} and \citet{Haywood08} argued that the large
scatter in the age-metallicity relation near the Sun
\citep[e.g.:][]{Edvardsson93, Bergemann14} is evidence of radial migration,
whereas \citet{Lee11}, using SEGUE data \citep[{\it Sloan Extension for
Galactic Understanding and Exploration},][]{Yanny09}, invoked radial
migration to explain why the metallicity of thin disc stars (for the range
$-0.5 \lesssim$[Fe/H]$\lesssim +0.2\dex$) is uncorrelated with their orbital
eccentricity. 
Finally, using RAVE data \citep[{\it RAdial Velocity Experiment},][]{Steinmetz06}, \citet{Minchev14}  found a decline of the velocity dispersion of the $\alpha-$enhanced low-metallicity disc stars, and suggested that the stars responsible for this decline are migrators from the inner disc.

The actual efficiency of radial migration, i.e., the maximum distance from
which a star can reach the Solar neighbourhood, has never been
observationally constrained. Indeed, this is a challenging task, since
distinguishing radially migrated stars from ones of the same metallicity that
were born locally requires either accurate ages (for example, SMR stars
having migrated from the Bulge region should be on average older than locally
born stars at the same metallicity) or knowing how the metallicity gradient
of the ISM has evolved.  In this paper, however, we aim to obtain a first
estimate of the radial migration efficiency by investigating the shape of the
metallicity distribution function of the metal-rich tail of the Solar
neighbourhood stars, in combination with a study of stellar orbits. For this
purpose, we use the kinematically unbiased spectroscopic catalogue of RAVE
\citep{Steinmetz06}, for which the latest data release
\citep[DR4,][]{Kordopatis13b} has published the atmospheric parameters,
metallicities and distances of approximately 400,000 relatively bright FGK
stars ($9<I<12$~mag).

Section~\ref{sect:data_description} describes the data set used, in
particular the new calibration of the metal-rich end, as well as the way the
distances, velocities and orbits of the stars have been computed.
Section~\ref{sect:characterisation} characterises the significance of the SMR
stars that we identify in RAVE, and shows that the normal disc giants is likely to have
amongst them stars born in regions where the Galactic bulge now dominates.
Finally, Section~\ref{sect:conclusions} concludes.

%%%%%%%%%%%%%%%%%%%%%%%%%%%%%%%%
\section{Description of the data and the new metallicity  calibration relation}
\label{sect:data_description}

\subsection{A new metallicity calibration for the metal-rich stars}

\begin{figure}
\centering
\includegraphics[width=0.99\linewidth, angle=0]{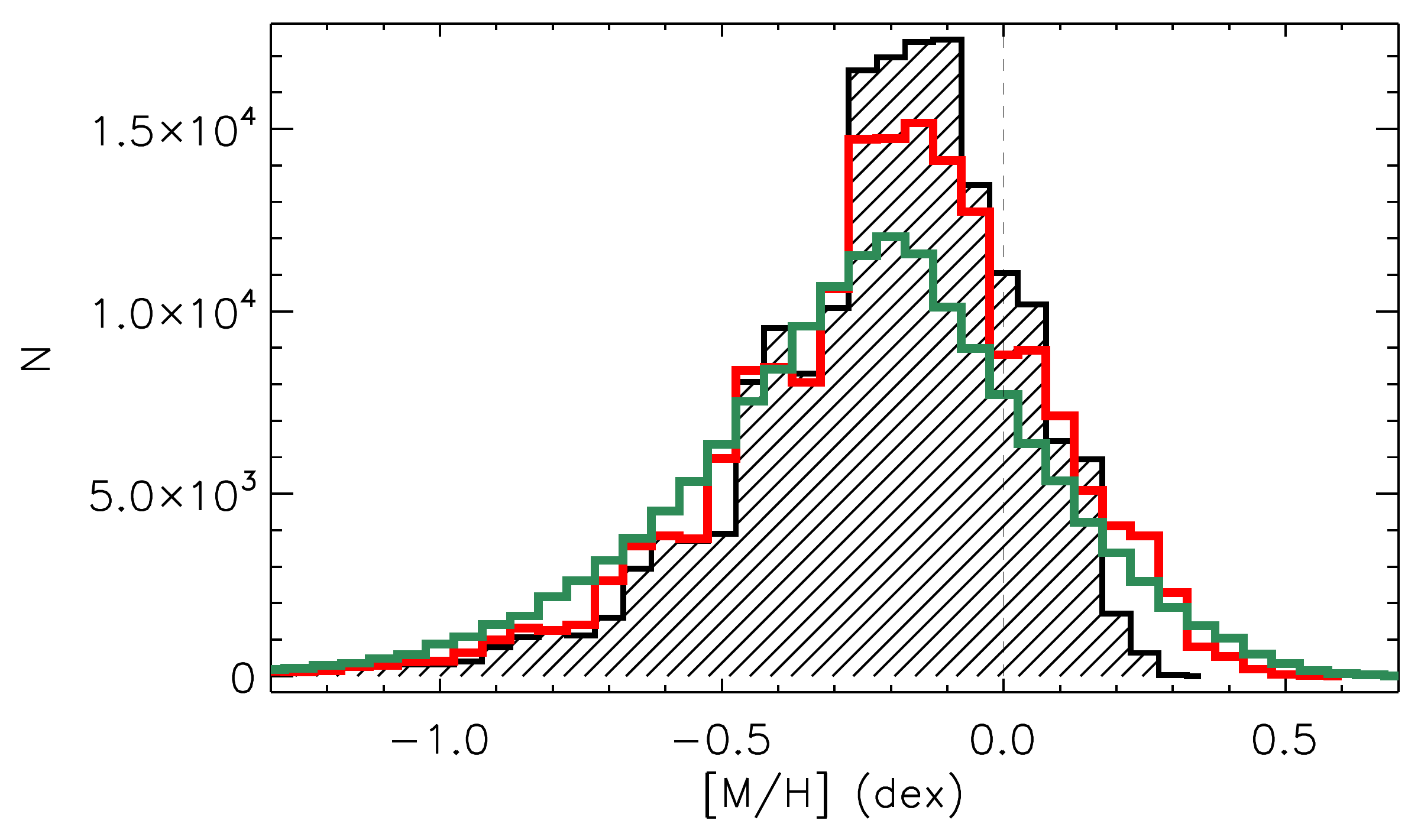}
\caption{Metallicity distributions of the selected RAVE quality  subsample.  The $x-$axis shows the DR4 metallicities (black histogram), the DR5 metallicities (red histogram), or the iron abundance ([Fe/H], green histogram) coming from the DR4 chemical pipeline. The $y-$axis shows the number of RAVE stars, N, in each bin. }
\label{fig:MDFs_methods}
\end{figure}

One of the major improvements of RAVE DR4 \citep{Kordopatis13b}, compared to
the previous data releases \citep{Zwitter08,Siebert11}, is its more thorough
metallicity calibration, based on the RAVE observations of cluster stars and
the availability of high-resolution spectra of already observed RAVE targets.
Although the calibration has had several successes
\citep[e.g.:][]{Kordopatis13c,Binney14a,Conrad14, Minchev14,Kordopatis14}, it suffered
from a lack of calibration targets at the high-metallicity end.
High-metallicity stars are not $\alpha$-rich, so at the high-metallicity end
the [M/H] and [Fe/H] distributions should approximately coincide.  In
Fig.~\ref{fig:MDFs_methods} the black histogram shows the DR4 [M/H]
distribution while the green histogram shows the [Fe/H] distribution from the
RAVE chemical pipeline \citep{Boeche13}. Contrary to expectation, the [M/H]
distribution falls far below the [Fe/H] distribution at the high-metallicity
end.

In the light of this discrepancy, the RAVE DR4 metallicity calibration has
been revised at the metal-rich end, using spectra of Gaia Benchmark stars
\citep{Jofre14} processed through the RAVE pipeline, as well as a comparison
of the pipeline's results for an additional 150 metal-rich stars that had
parameters derived from very high resolution spectra using the {\it High Accuracy Radial velocity Planet Searcher} (HARPS) spectrograph
%HARPS
\citep{Adibekyan13} and the {\it  Fiber-fed Extended Range Optical Spectrograph} \citep[FEROS][]{Worley12}.
Details of the updated calibration will be given in the RAVE-DR5 paper (in
prep.). In summary, the calibration procedure is the same as in DR4, i.e.,
fitting the difference between the metallicities derived from the pipeline
and literature metallicities of all available calibrators to a second order
polynomial in \logg~ and \meta.  The resulting calibration relation is almost
unchanged for all of the stars with $\meta\lesssim0$ (less than $\sim
0.05\dex$ difference), and provides a more symmetric shape of the metallicity
distribution function at $\meta>0$ -- the red histogram of
Fig.~\ref{fig:MDFs_methods} shows the new distribution. 

\subsection{Distances, positions, orbits and quality sub-sample}

\begin{figure}
\centering
\includegraphics[width=0.7\linewidth, angle=-90]{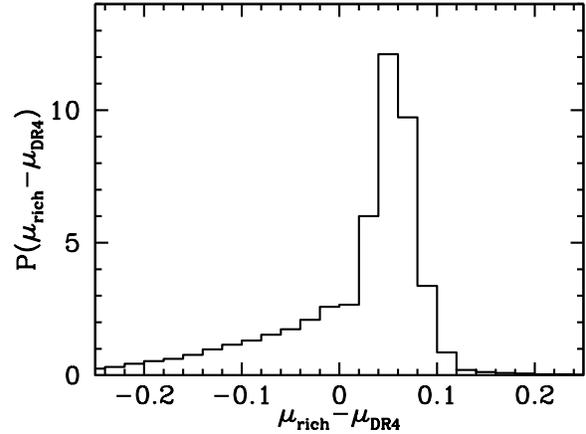}
\caption{Probability density ($1/\mu$) of the change in the distance modulus of the metal-rich stars when their DR4 metallicity is increased by 0.15~dex, which is the maximum effect that DR5 metallicity calibration has on DR4 metallicities. The mean effect is an increase of 0.05 in the distance modulus, which translates into an increase of about 2\% in the derived line-of-sight distances. }
\label{fig:distance_shift}
\end{figure}

After modifying the metallicities of the stars, one should re-determine the
distances to these stars since the DR4 distances \citep{Binney14a} used the
DR4 metallicities. However, by re-running the distance pipeline it has
been found that increasing the metallicities of all super-solar metallicity
stars by up to $+0.15\dex$ typically adds $\sim0.05$ to the derived distance
moduli (see Fig.~\ref{fig:distance_shift}). This represents about a 2\%
increase in distance, which is negligible compared to our uncertainties,
estimated to be $\sim15$\%. Hence here we use the published DR4 distances.  

We adopted the solar motion with respect to the Local Standard of Rest (LSR)
of \cite{Schonrich10}, namely $U_0=11.1, V_0=12.24,
W_0=7.25\kms$, and assumed that the Sun is located at
$(R_0,Z_0)=(8,0)\kpc$ and that the LSR is on a circular orbit with
circular speed $V_c=220\kms$. Then from the DR4 data we computed the
Galactocentric positions and velocities of the stars, in the same way as in
\citet{Kordopatis13c}.

We used the model Galaxy of \citet{Dehnen98a}, where the Galactic
gravitational potential is built with three superposed double-exponential
discs (thin disc, thick disc and gas layer) and two spheroids (bulge and dark
halo). More specifically, the density of each disc is given by:

 \begin{equation}
\rho(R,z)=\frac{\Sigma_0}{2z_{\rm d}}\exp\left[-\left( \frac{R_{\rm h}}{R} + \frac{R}{R_{\rm d}} + \frac{|z|}{z_{\rm d}} \right)\right],
 \end{equation}
 where $R$ and $z$ are the coordinates in a Galactocentric cylindrical
coordinate system, $R_{\rm d}$ and $z_{\rm d}$ are, respectively, the
scale-length and scale-height, $\Sigma_0$ is the disc's central surface
density and where a non-zero value of $R_{\rm h}$ generates a central
depression in the disc.  The density of the spheroids is given by:
  \begin{equation}
\rho(R,z)=\frac{\rho_0}{m^\gamma(a+m)^{\beta-\gamma}}\exp[-(m r_0 / r_{\rm cut})^2], 
 \end{equation}
 where $\beta$ and $\gamma$ control the outer and inner density slopes, $r_0$
and $r_{\rm cut}$ are the scale and cut-off radii, $\rho_0$ sets the scale
density and $m$, defined as:
   \begin{equation}
m(R,z) \equiv \sqrt{(R/r_0)^2 + (z/qr_0)^2},
 \end{equation}
 includes the axis ratio $q$ of the isodensity surfaces.
Table~\ref{tab:potential} presents the values that are adopted in this work
for each disc and spheroid. In the fixed potential of this model we used the
``St\"ackel Fudge'' of \cite{Binney12a} to determine the smallest $\rp$ and
largest $\ra$ radii at which the orbit defined by its given initial
condition cuts the Galactic plane, and then computed the orbit's mean radius
$\rb\equiv(\rp+\ra)/2$ and the orbital eccentricity { $e\equiv(\ra-\rp)/(\ra+\rp)$.} 
 
\begin{table}%[htdp]
\caption{Parameters for the adopted mass model of the Milky Way.}
\label{tab:potential}
\begin{center}
\begin{tabular}{cccc}
\hline \hline
Disc & Thick & Thin & Gas \\\\

$\Sigma_0{\,({\rm M_\odot}\kpc^{-2})}$ & $7.30\times10^7$ & $1.11\times 10^9$ &$1.14\times10^8$ \\
$R_{\rm d}~(\kpc)$ & 2.4 & 2.4 & 4.8 \\
$z_{\rm d}~(\kpc)$ & 1.0 & 0.36 & 0.04 \\
$R_{\rm h}~(\kpc)$ & 0 & 0 & 4\\ \\

Spheroid &  Dark halo & Bulge\\
$\rho_0{\,({\rm M_\odot}\kpc^{-3})}$ & $1.26\times 10^9$ &  $7.56\times10^8$ \\
$q$ & 0.8 & 0.6 \\
$\gamma$ & $-2$ & 1.8 \\
$\beta$& 2.21 & 1.8 \\
$r_0~(\kpc)$ & 1.09 & 1\\
$r_{\rm cut}~(\kpc)$& 1000 & 1.9 \\
\hline
\end{tabular}
\end{center}
\label{default}
\end{table}%

The sample analysed here is selected to have reliable stellar parameters (and
therefore distances, velocities and orbits) following the recommendations of
\citet{Kordopatis13b}. It contains only stars that have effective temperature
\teff$> 4000\,$K, surface gravity $\log g>0.5\dex$, errors in line-of-sight
velocity e$V_\parallel<10\kms$, spectral morphological flags set to `$n$'
(normal stars) and for which the stellar parameter pipeline had
converged\footnote{Given the S/N threshold applied, $algo\_conv=0$ was
required, indicating that the pipeline should converge without getting
outside the synthetic spectra grid boundaries \citep[see also][]{Kordopatis11a}.}. In addition, we selected,
for higher accuracy, only stellar parameters obtained from spectra with
a Signal-to-Noise ratio (S/N) higher than 20.

%%%%%%%%%%%%%%%%%%%%%%%%%%%%%%%%
\section{Characterisation of the metal-rich population}
\label{sect:characterisation}
\subsection{Identification of the strength of the signal}

 \begin{figure}
\centering
\includegraphics[width=\linewidth, angle=0]{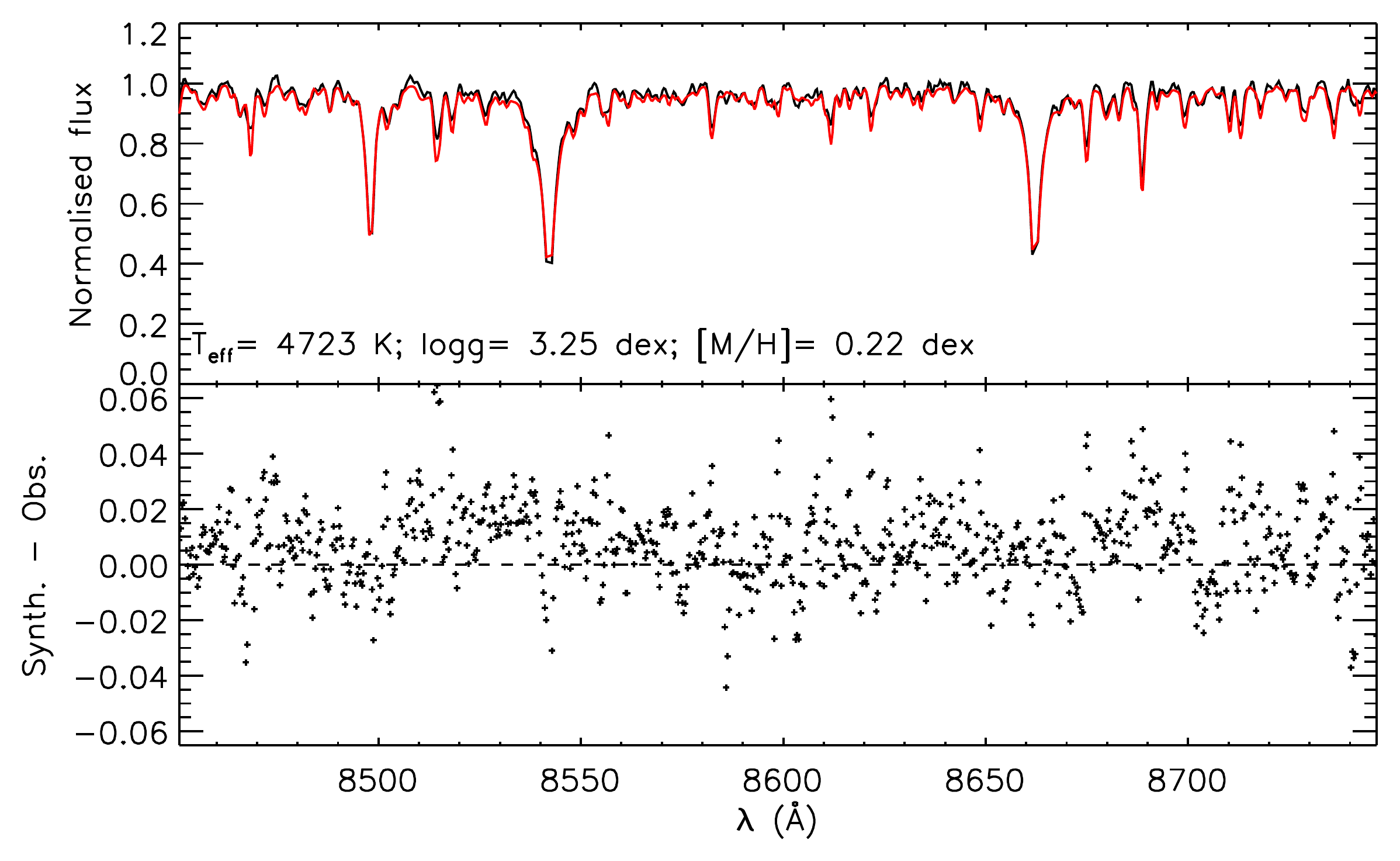}
\caption{{\bf Top:} Observed (in black) and synthetic (in red) spectra of a RAVE super metal-rich star. {\bf Bottom:} Residuals between the observations and the synthetic spectrum.}
\label{fig:spectrum}
\end{figure}

\begin{figure}
\centering
$\begin{array}{cc}
\includegraphics[width=0.5\linewidth, angle=0]{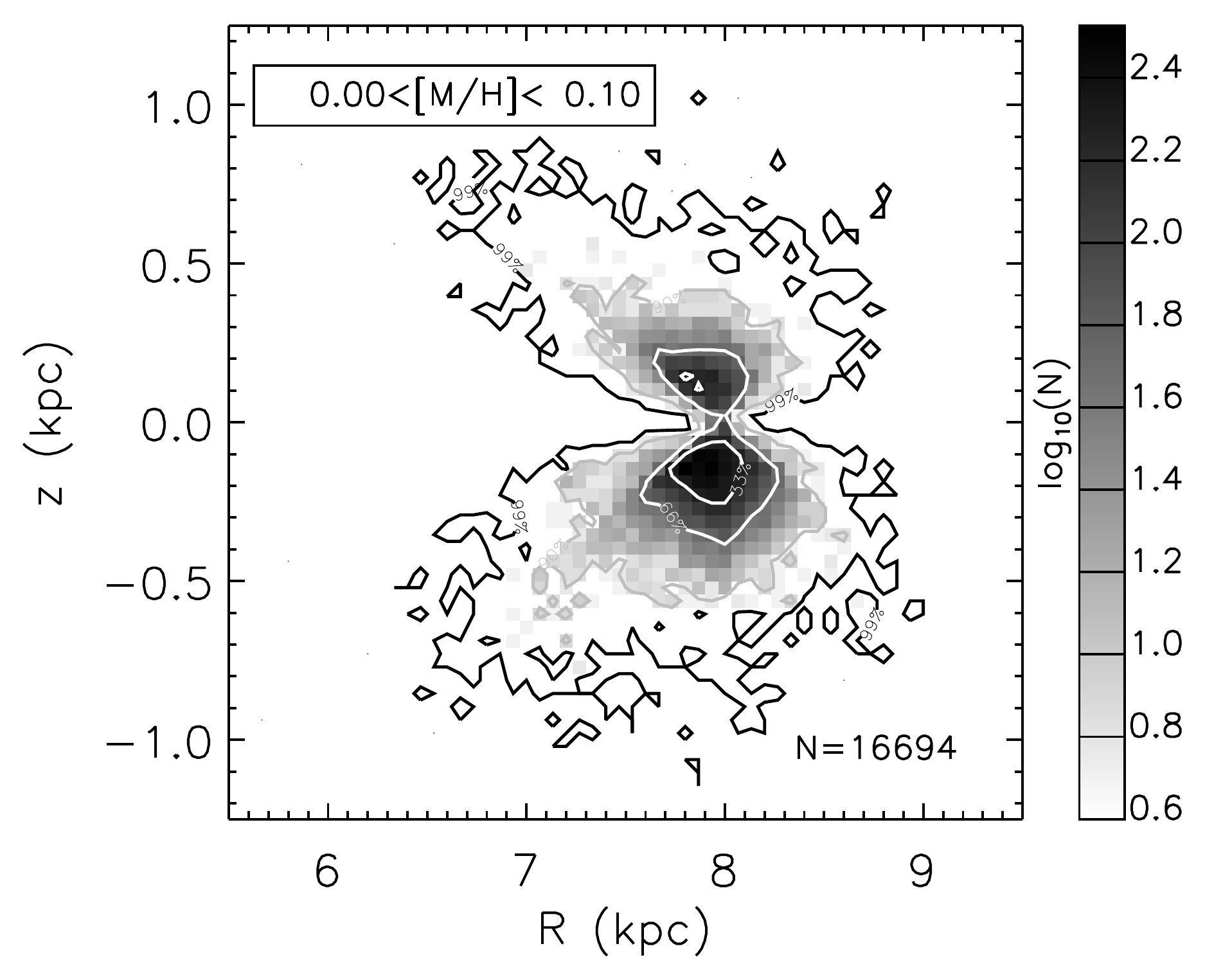}& 
\includegraphics[width=0.5\linewidth, angle=0]{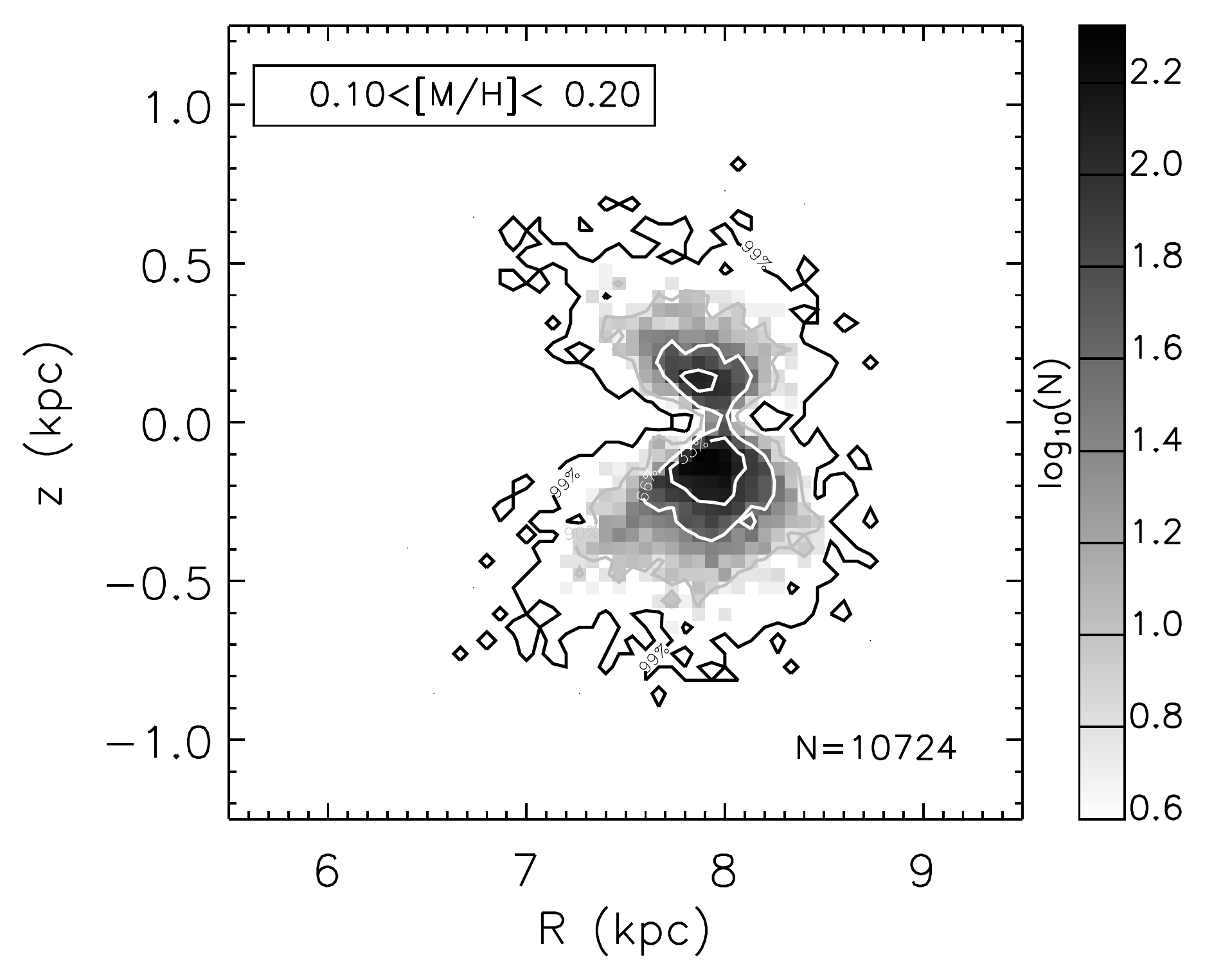} \\
\includegraphics[width=0.5\linewidth, angle=0]{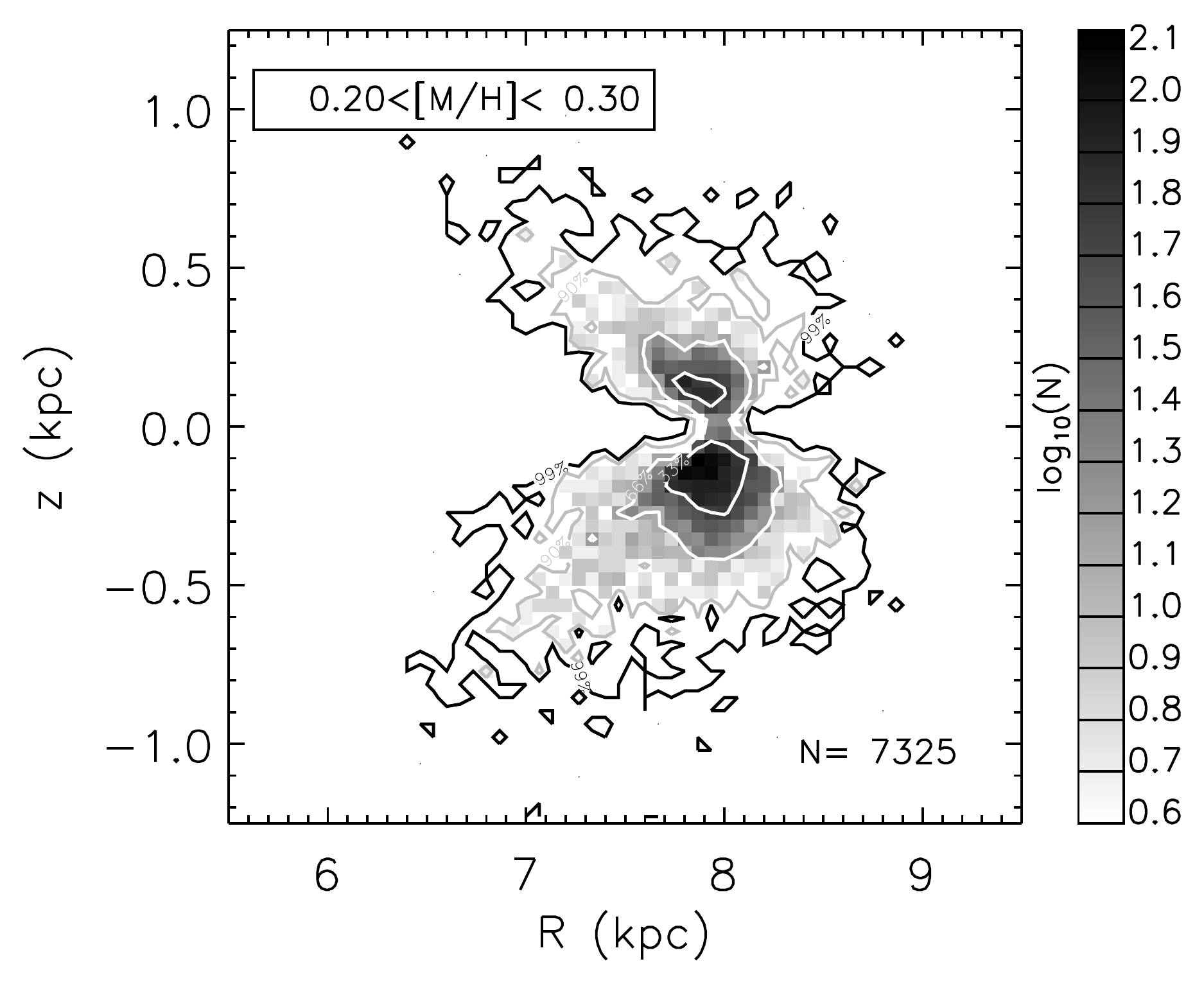}&
\includegraphics[width=0.5\linewidth, angle=0]{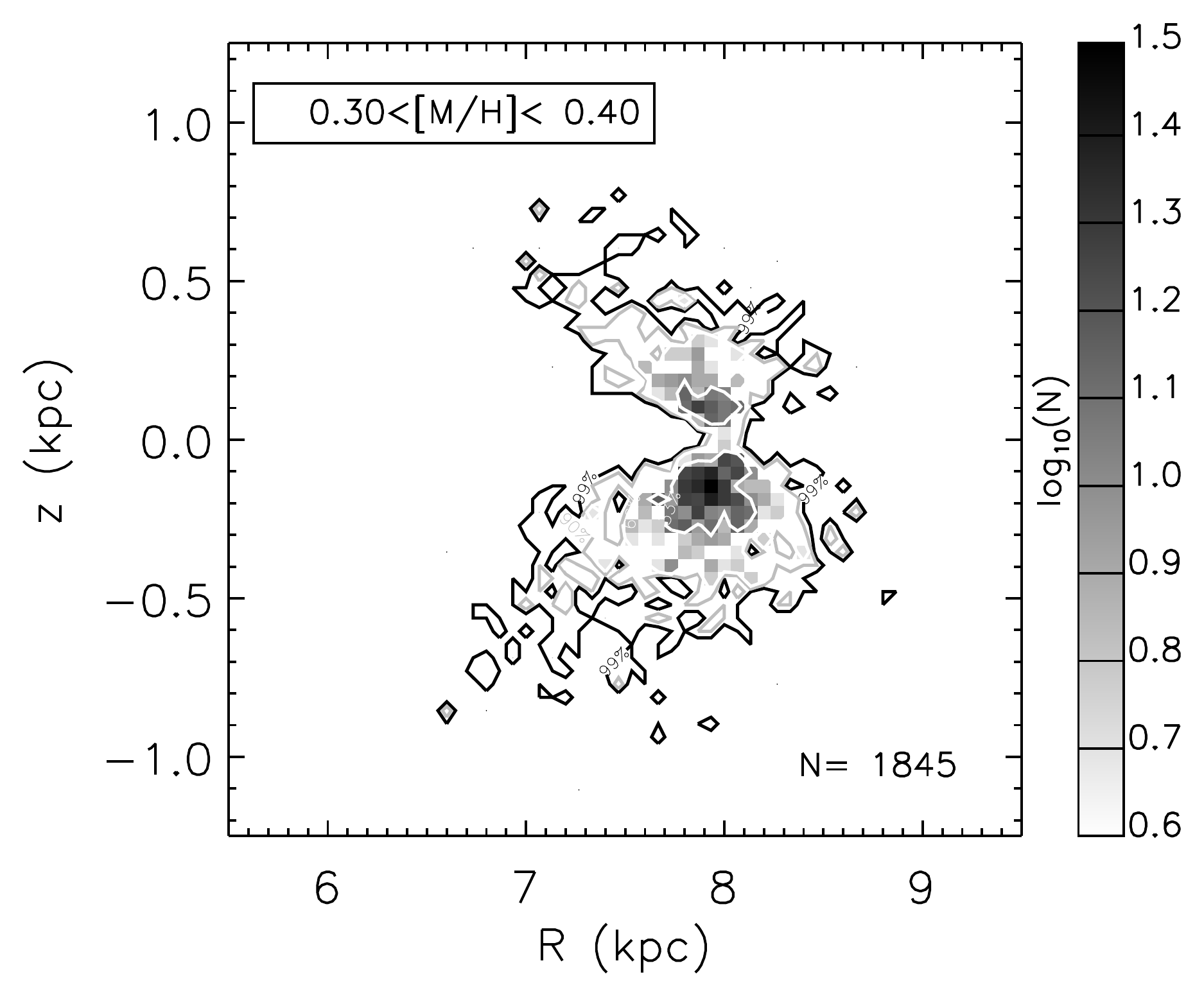}
\end{array}$
\caption{Positions of the metal-rich stars (in bins of $0.1\dex$) in the $(R,z)$ plane. The grey-scale is the logarithmic number of the targets. Most of  targets are confined within $0.5\kpc$ of  the Galactic plane, with however a non negligible number reaching distances up to $1\kpc$.}
\label{fig:metal_rich_map}
\end{figure}

Figure~\ref{fig:MDFs_methods} shows the distributions of the DR4 metallicity
(black points) and iron (green points) abundances for the selected quality
sample; the  red points show the distribution obtained with the new
calibration (DR5). One can see the effect of the new DR5 calibration for the
super-Solar metallicity stars: they are now in better agreement with [Fe/H]
in the shape of the tail of the distribution.  In particular, one can notice
that the selected RAVE quality subsample has more than $4\times10^4$ stars
with $\meta>0$. A visual inspection of the fit of the synthetic templates
to the observed spectra confirmed that the results were in good agreement with
the data (Fig.~\ref{fig:spectrum}), and that within the errors we could trust
the derived parameters \citep[see][for a discussion on the internal errors of the method]{Kordopatis13b}.  These targets, all of which have $\log g\geq 2\dex$,
are located relatively close to the Sun ($R=8\pm 1.5\kpc$,
Fig.~\ref{fig:metal_rich_map}), and are mainly located near the Galactic
plane, with nevertheless, some stars seen up to $1\kpc$.

\begin{figure}
\centering
\includegraphics[width=0.99\linewidth, angle=0]{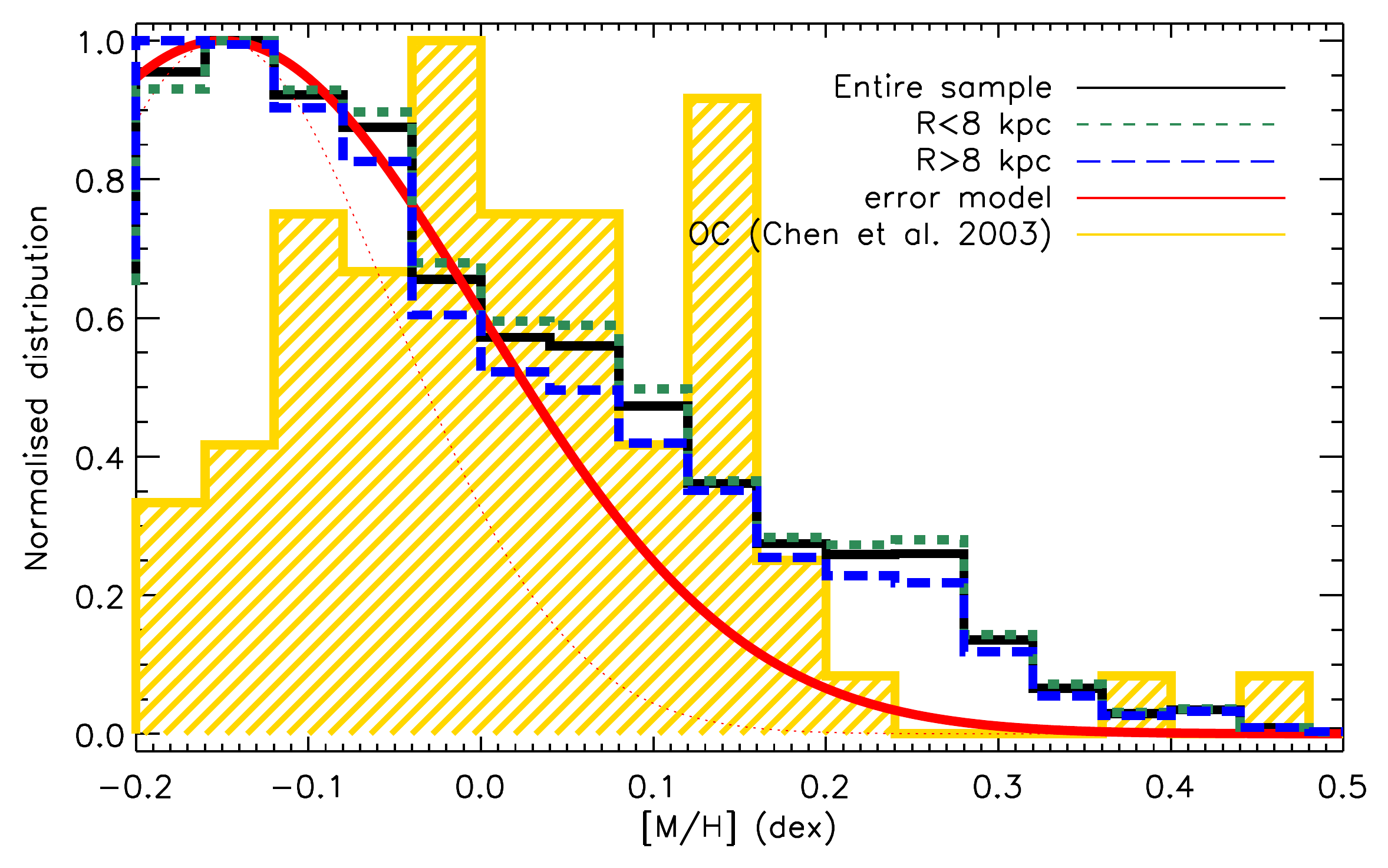}
\caption{Normalised metallicity distribution function (MDF) of the metal-rich stars for the entire quality sample (black), the inner Galaxy one (green) and outer Galaxy one (blue). The MDF of 82 open clusters \citep[out of 118 published by ][]{Chen03}, in the distance range $7.5<R<10\kpc$,  is also illustrated in yellow. Finally, the red Gaussians have standard deviations of 0.1 (dashed line) and $0.15\dex$ (solid line) and represent the published and an over-estimated internal error of the DR4 metallicity determination, respectively. }
\label{fig:MDFs_Radii}
\end{figure}

Figure~\ref{fig:MDFs_Radii} shows that the metallicity distribution function
has roughly the same shape in the inner ($R<8\kpc$) and outer
($R>8\kpc$) Galaxy, although the inner Galaxy is marginally more
metal-rich.  However, we now ask to what extent the number of stars measured
to be metal-rich is boosted by stars scattered by observational error from
the sub-solar peak of the metallicity distribution ($\meta\sim -0.15\dex$, see
Fig.~\ref{fig:MDFs_methods}).  Suppose the uncertainty of the sub-Solar
metallicity stars follows a Gaussian distribution, of standard deviation
$0.1\dex$, as suggested in Tables~1~and~2 of \citet{Kordopatis13b}.  Then the
red dotted Gaussian of Fig.~\ref{fig:MDFs_Radii} indicates that only a small
fraction of the stars with measured $\meta>0.2\dex$  would be generated by
accidental scattering of sub-solar stars. Indeed, the number of stars seen at
$\meta>0.05\dex$ cannot be explained by accidental scattering even when the
errors are over-estimated by 50\% by setting $\sigma=0.15\dex$ (red solid
Gaussian in Fig.~\ref{fig:MDFs_Radii}). 

To further assess the plausibility of metal-rich stars really existing, in
Fig.~\ref{fig:MDFs_Radii} we compare our [M/H] distribution with the [Fe/H]
distribution of the 82 open clusters published in \citet{Chen03} that have
Galactocentric radii in the range probed by RAVE {\bf ($7.5<R<10\kpc$, yellow
histogram)}.  The metallicity distribution of the clusters is narrower than
that of the stars, having its principal peak at $\meta\simeq0$ rather than
$\meta\simeq-0.17\dex$, and, with the exception of two outliers, falling to
zero at $\meta=0.2\dex$. The paucity of clusters with $\meta<0$ is a natural
consequence of the monotonic increase with time in the ISM's metallicity and
the youth of clusters -- the clusters are in the majority younger than
$1.5\Gyr$ \citep{Chen03} whereas the RAVE metal-rich stars should be a few
billion years old given the typical age ($\sim 5\Gyr$) of Solar-metallicity
field stars.  The presence of old field stars more metal-rich than any but
the two outlying clusters would be hard to explain in the absence of radial
migration because the cluster distribution implies that even now, and more so
in the past, in the probed radial range the gas is too metal-poor to form these
stars.

\bigskip

\begin{figure}
\centering \includegraphics[width=0.79\linewidth, angle=0]{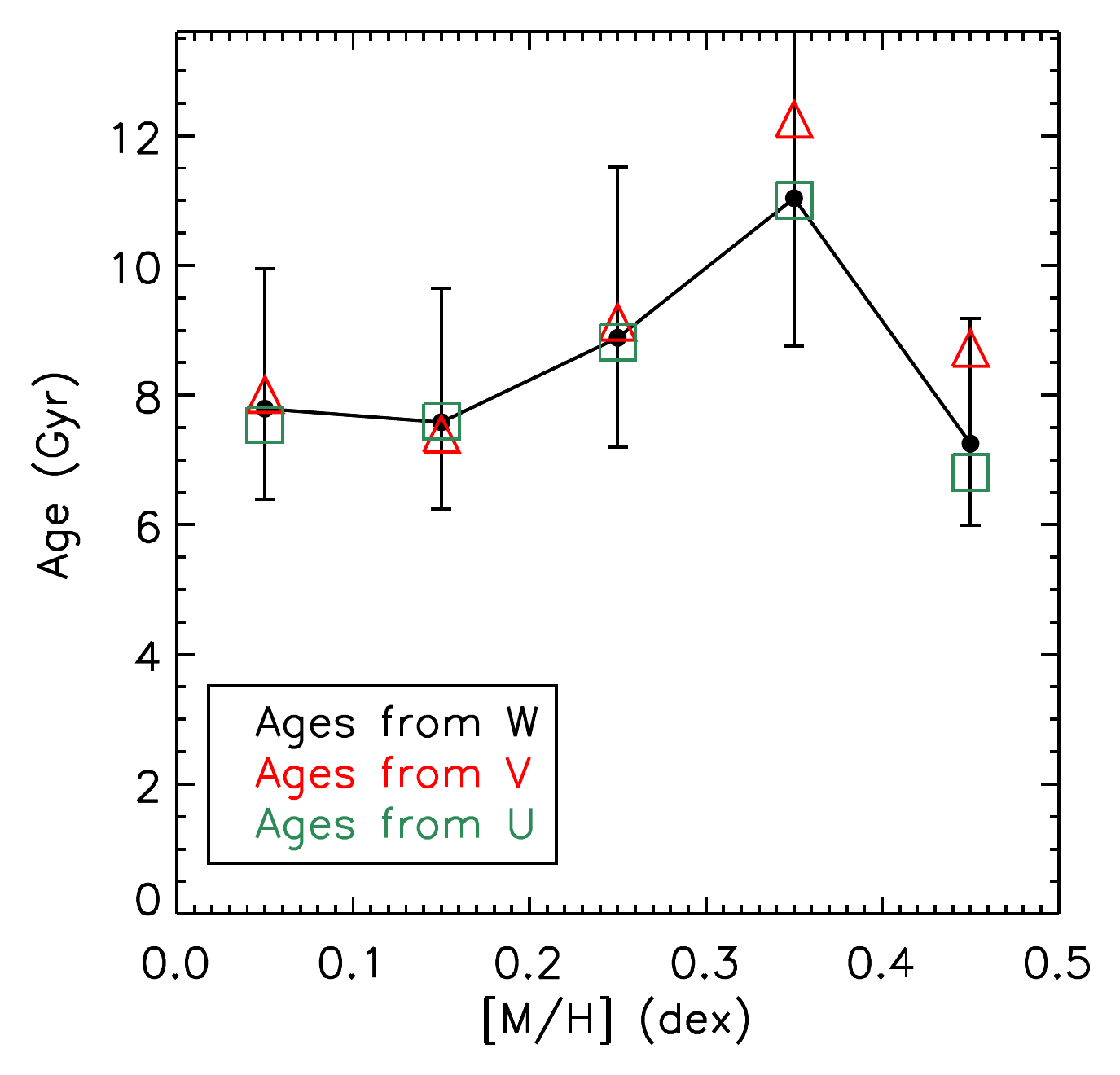}
\caption{Age estimation of the metal-rich stars using the age-velocity
dispersion relations for the $\rm U,V,W$ heliocentric velocity components. All age estimators are
consistent with these being old stars. } \label{fig:Ages}
\end{figure}

We obtained a \emph{rough} estimation of the age of the RAVE stars by
applying, for different metallicity bins, the age--velocity dispersion
relation defined for the heliocentric $\rm U,V,W$ velocities as:
 \begin{equation}\label{eq:age_vel}
 \sigma=a\times \hbox{age}^{k},
 \end{equation}
 where $\sigma$ is the velocity dispersion ($\rm \sigma_U,\sigma_V$ or
$\rm \sigma_W$), $a$ is the local normalisation factor, set in order to be equal
to the velocity dispersion of $1\Gyr$ old stars, and $k$ is is power-law
exponent defining the age-velocity dispersion law.  By adopting the values
$(a_U, a_V, a_W)=(27, 18, 10)\kms$ \citep{Robin03} and $(k_U, k_V,
k_W)=(0.31, 0.34, 0.47 \pm 0.05)$ \citep{Nordstrom04}, we confirm the
previous statement that the metal-rich stars in our sample are on average
old  (see Fig.~\ref{fig:Ages}), with approximate ages $\sim8\pm2\Gyr$.
{We note that the age uncertainties are expected to be larger, as they} 
% We have not
%corrected the velocity dispersions for the measurement errors, as these ages
%are only indicative, and the ages one derives do 
depend on the values adopted
for the parameters $a$ and $k$ in equation (\ref{eq:age_vel}) \citep[see
Table~8 of ][for a review of the possible values of $k$]{Sharma14}. Our
argument is simply that if the stars under study were young, their velocity
dispersions would be significantly smaller than they actually are.

%%%%%%%%%%%%%%%%%%%%%%%%%%%%%%%%%%%%
\subsection{Identification of the radially migrated stars}
\label{sec:radial_migration_candidates}

\begin{figure}
\centering
\includegraphics[width=0.99\linewidth, angle=0]{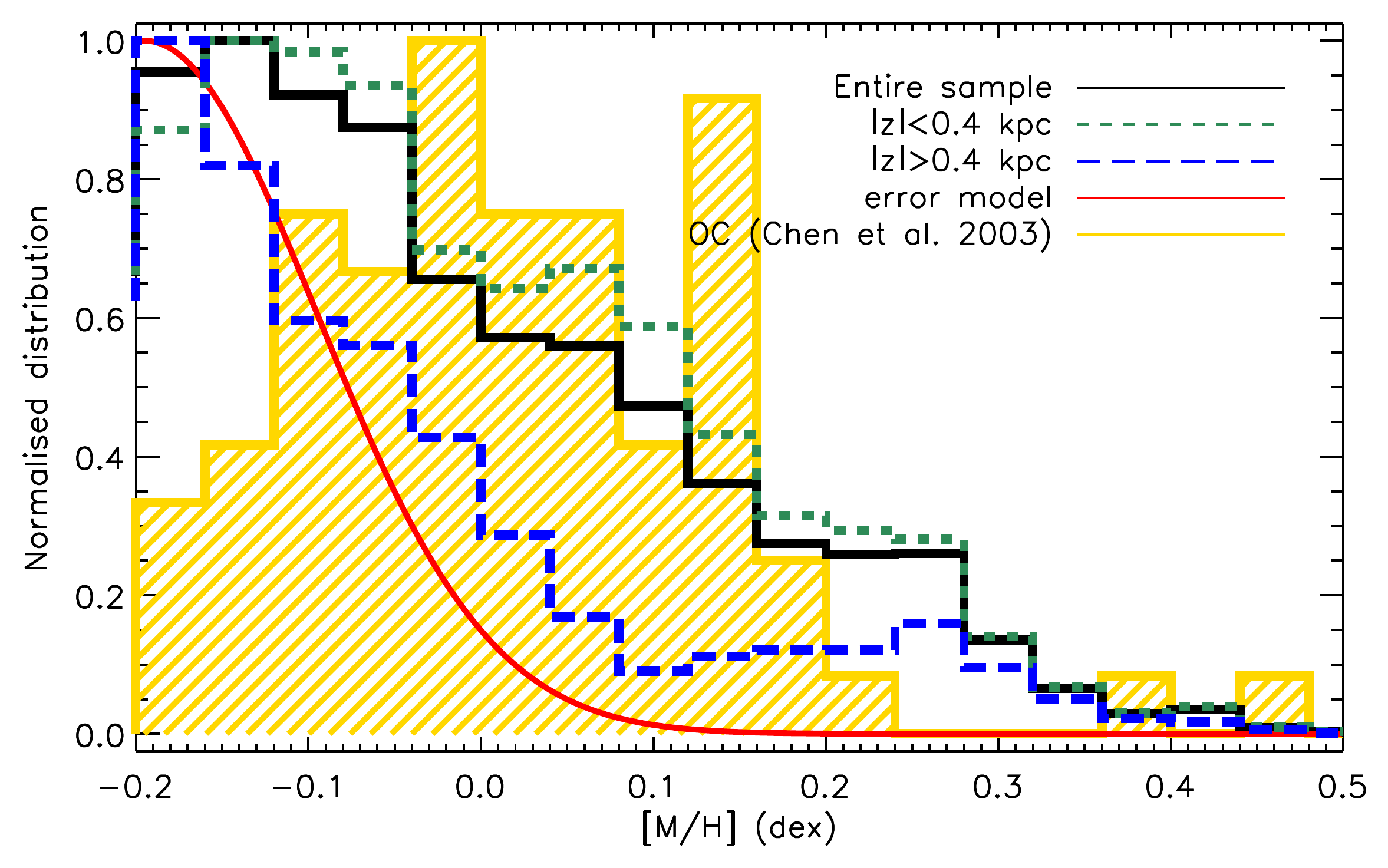}
 \caption{Same as Fig.~\ref{fig:MDFs_Radii}, but for a separation as a
function of the distance from the Galactic plane. The error model is a
Gaussian of $\sigma=0.1\dex$, centred at the peak of the blue histogram, at
$\meta=-0.2\dex$. } \label{fig:MDFs_Z01}
\end{figure}

As division of the sample by Galactocentric radius did not reveal any
significant difference in the shape of the metallicity distribution
functions, in Fig.~\ref{fig:MDFs_Z01} we split our sample into closer and
farther than $0.4\kpc$ from the plane\footnote{Other threshold values have
been tested, up to $0.8\kpc$ in order to have enough stars in each
sub-sample, without changing our conclusions.} \citep[note that given the
RAVE footprint on the sky, stars far from the plane are preferentially
located towards the inner Galaxy, see Fig.~\ref{fig:metal_rich_map}
and][]{Kordopatis13b}. 

The
metallicity distributions of Fig.~\ref{fig:MDFs_Z01} indicate that
an uncertainty of $0.15\dex$ in metallicity, as assumed above, is indeed an
over-estimation of the internal errors, since the high-altitude sample (blue
broken curve) shows a narrower distribution in $\meta$. As discussed in the
previous section, an uncertainty of $0.05$ to $0.1\dex$ is more realistic. 

Whereas the metallicity distribution of stars near the plane (green
histogram) falls smoothly from a peak at $\meta\sim-0.125\dex$ to near zero at
$\meta\sim0.5\dex$, that of stars farther than $0.4\kpc$ from the plane (blue histogram)
falls smoothly from a peak at $\meta\sim-0.175\dex$ to a local minimum at
$\meta\sim0.1\dex$ and then flattens to a very extensive tail. The tail rises to
a peak around $\meta=0.275\dex$ and then gradually fades. From $\meta=0.15\dex$ the
tail comprises the SMR stars.

\begin{figure}
\centering \includegraphics[width=0.99\linewidth, angle=0]{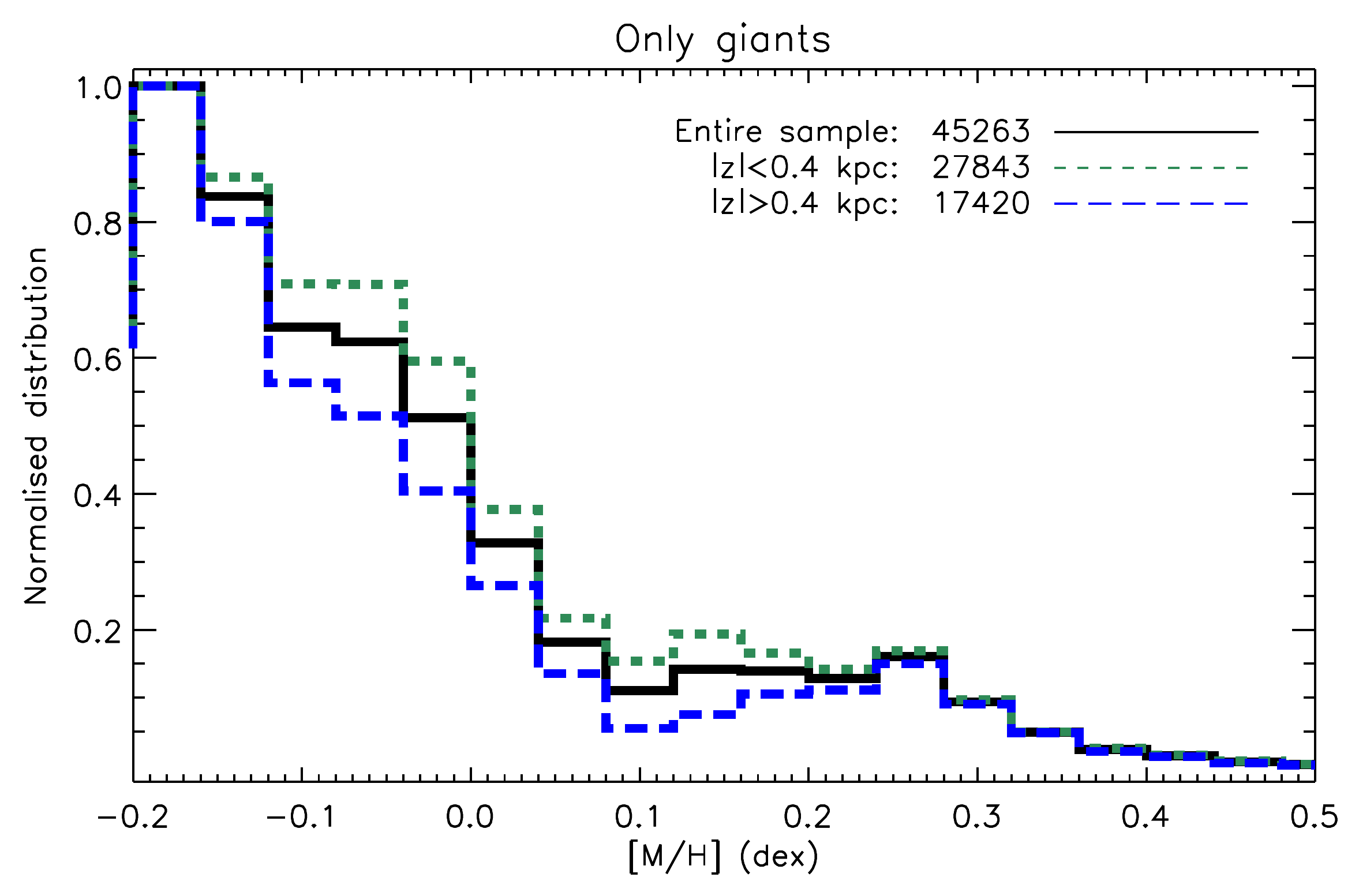}
\caption{Same as Fig.~\ref{fig:MDFs_Z01}, but only for stars with
$\log g<3.5\dex$. For $\meta>0.1\dex$ the metallicity distribution is flat for
both the subsamples: that close to the plane and that far from it.}
\label{fig:MDFs_Z03}
\end{figure}

In Fig.~\ref{fig:MDFs_Z03} the histograms show the metallicity distributions
of just the giants ($\log g<3.5\dex$): the green histogram is for those that lie
closer than $0.4\kpc$ to the plane, the blue histogram for those further than
$0.4\kpc$ from the plane and the black histogram is for the joint sample. The
blue histogram differs little from that shown in Fig.~\ref{fig:MDFs_Z01}
because all the contributing stars are quite distant and are unlikely to make
the magnitude cut if they are not giants. But the sample of stars that are
close to the plane is substantially modified by restricting the sample to
giants, and we see that the metallicity distribution of the giants is
essentially the same near and far from the plane. That is, in
Fig.~\ref{fig:MDFs_Z01} the difference between the blue and black histograms
is attributable by the contribution of the dwarfs. Many of the dwarfs are
younger than the giants,\footnote{Only few young giants are expected to be
present in our selection, due to their relatively short lifetime in that
evolutionary phase.} so they tend to be more metal-rich.
Figure~\ref{fig:MDFs_Z03} reveals that the metallicity distribution of the
in-plane giants has a long tail to match that of the giants that are further
away. Of course this is physically essential because most of the giants that
are currently near the plane will in $\sim50\Myr$ be far from it, and vice
versa. 

The key finding is that the (near-plane)   dwarfs {do not contribute
significantly above $0.25\dex$ even though they are typically younger. If these
SMR stars formed near us, we would expect them to have formed recently and
include RAVE dwarfs. This is clearly not the case, at least for $\meta \gtrsim 0.25\dex$.

The natural explanation of the tail is that it is made up of stars that have
migrated to us from smaller radii, where high metallicities were
achieved very early on \citep[e.g., Fig.~3 of][]{Minchev13}.  
 The distance that  the stars on the metal-rich end of the tail  need to have travelled to match the observations,
depends on the rate
at which the disc enriched its metallicity at each radius over time \citep[see for example][]{Wyse89,Chiappini97}. 
 The likelihood of an individual  star reaching the Solar neighbourhood decreases with
distance to travel. However, the exponential rise inwards in the number of
stars available to make the journey will to an extent compensate for the
decreasing probability of coming far, with the result that the number in the
tail decreases remarkably slowly with increasing $\meta$. 
Future work will include developing such a model and investigating its plausibility in the light of the results of this study.

\begin{figure*}
\centering
\includegraphics[width=\linewidth, angle=0]{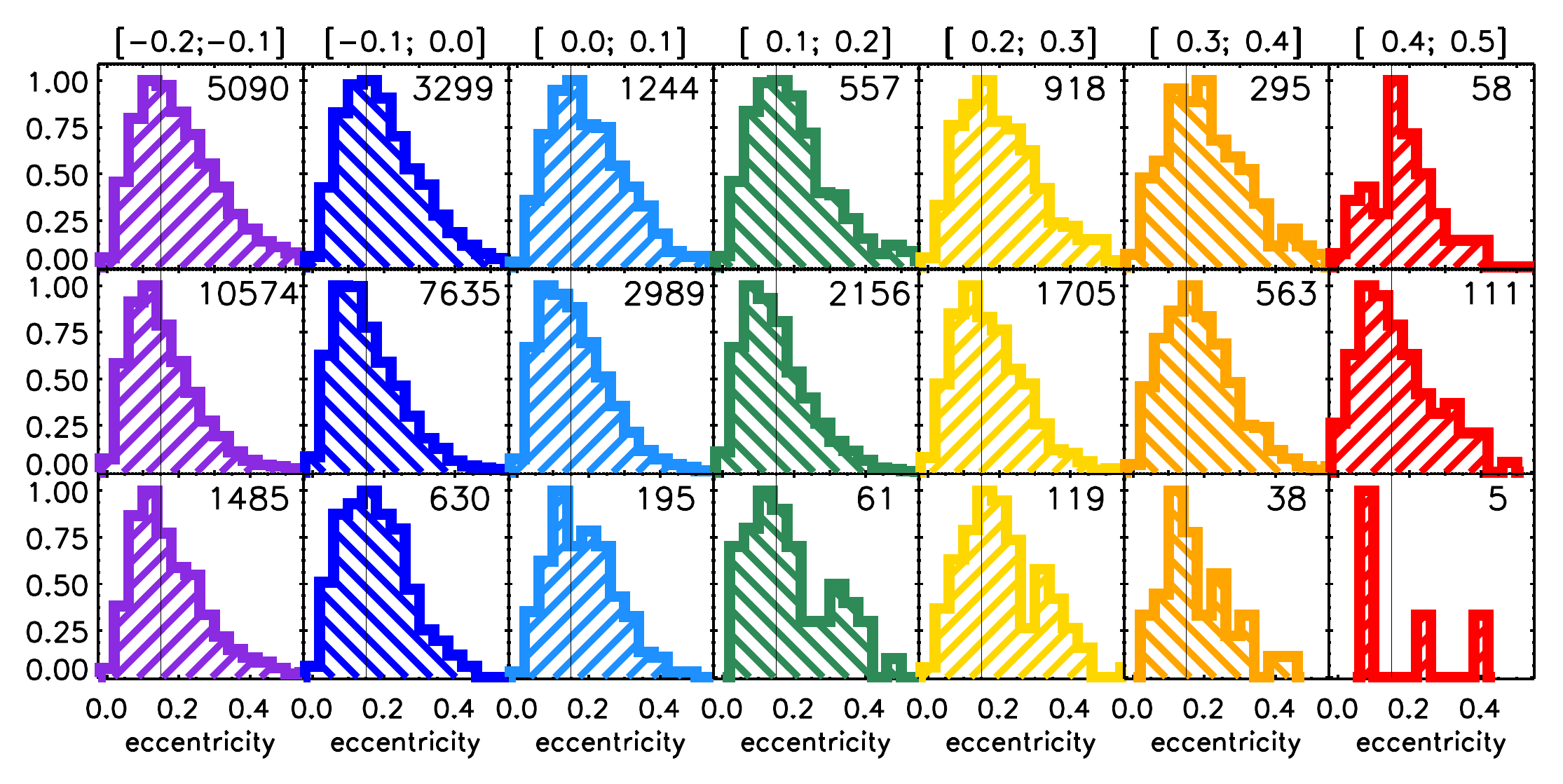}
 \caption{Eccentricity distribution (normalised to 1) of the RAVE giants
quality subsample in the inner Galaxy (6.5$<R<7.5\kpc$, top), in the Solar
neighbourhood (7.5$<R<8.5\kpc$, middle) and in the outer Galaxy (8.5$<R<9.5\kpc$,
bottom).  The histograms are obtained for stars of increasing metallicity, in
$0.1\dex$ wide bins, starting from $-0.2\dex$. For each histogram, the total
number of stars considered is noted in the upper right corner (the histograms
have been truncated to $e=0.55$). The plain vertical line in each panel is at
$e=0.15$, below which a star is considered having a circular orbit. The median and interquartile values of the ditributions are reported in Table~\ref{tab:interquartiles}.}
\label{fig:eccentricities}
\end{figure*}

\begin{figure*}
\centering
\includegraphics[width=\linewidth, angle=0]{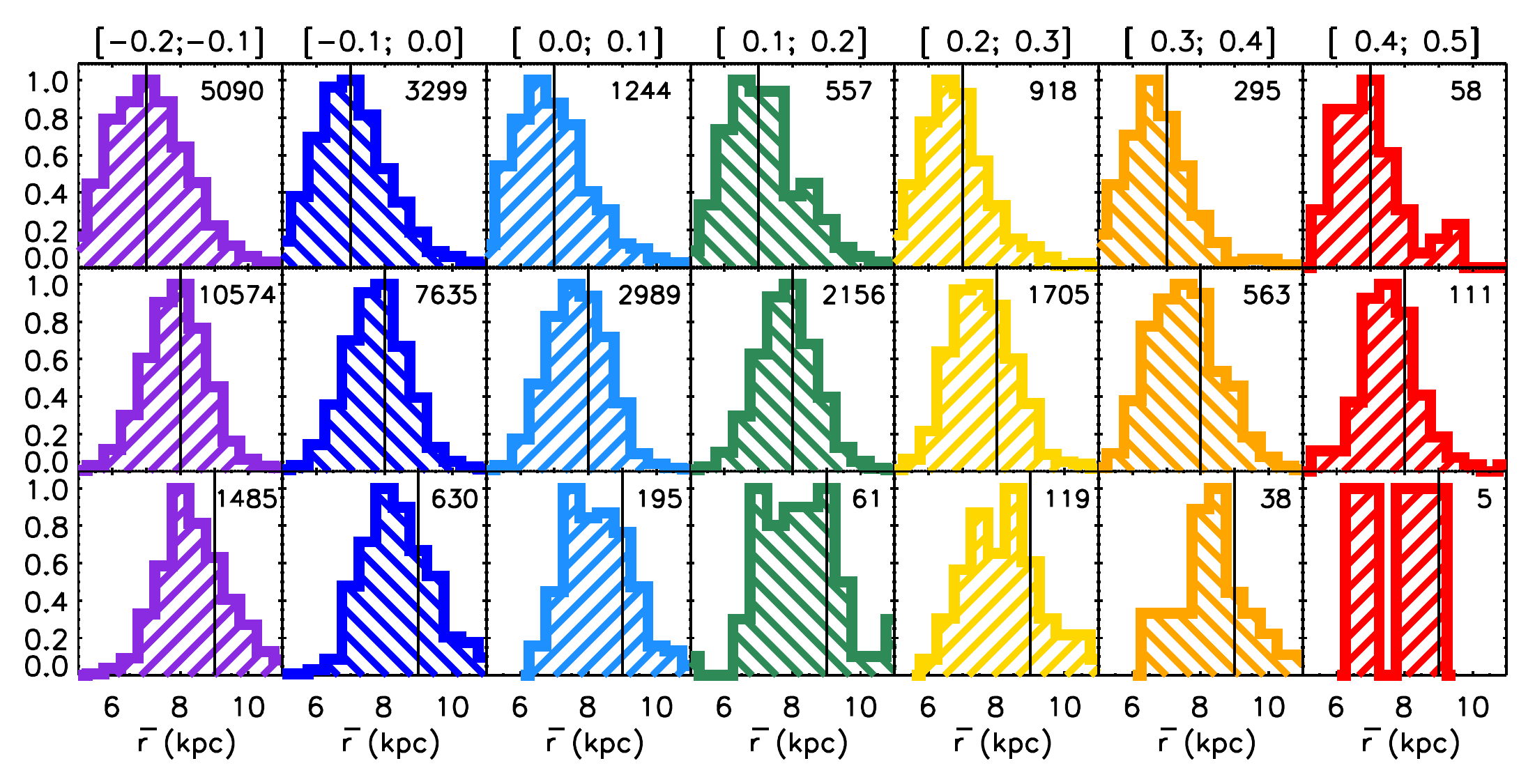}
\caption{Same as for Fig.~\ref{fig:eccentricities}, but for the distribution of the mean orbital radius ($\rb$) of the stars. In each panel,  the plain vertical line is at the mean observed radius. Histograms have been truncated at $[5,11]\kpc$.  }
\label{fig:mean_radii}
\end{figure*}

%%%%%%%%%%%%%%%%%%%%%%%%%%%%%%%%%
\subsection{Orbits of metal-rich stars}

We now examine the orbits of the metal-rich stars.
Figure~\ref{fig:eccentricities}  and Table~\ref{tab:interquartiles} illustrate and quantify the shape of the eccentricity distribution of the stars for three different distance ranges from the Galactic centre and for different metallicity bins, starting from $\meta\geq-0.2\dex$. Figure~\ref{fig:mean_radii} shows the mean orbital radii for the same stars.  From these two figures, one can see that:
\begin{itemize}
\item
 most of the stars have eccentricities
below $e\approx0.3$, with a peak at around $e\sim0.15$ indicating, as
expected for thin disc targets, that the stars are on nearly circular orbits.

\item
No systematic variation is evident of the eccentricity distribution with
metallicity. In particular, SMR stars are not more likely that
low-metallicity stars to be on eccentric orbits.

\item There is a slight tendency for stars to be observed nearer apocentre
than pericentre ($R>\rb$).  This bias towards apocentre may increase slightly
with metallicity, but the effect is at best weak. Given that the density of
stars decreases exponentially with guiding-centre radius, it is inevitable
that when we focus on nearby stars with large eccentricities, stars with
small guiding centres outnumber stars with large guiding centres, so the
majority of stars with high eccentricities will be seen nearer apocentre than
pericentre.  Indeed most of the solar-neighbourhood stars with $e \geq 0.3$
prove to have mean orbital radii $\rb \lesssim 6\kpc$.  Thus ``blurring'' by
eccentric orbits that have been populated by scattering at Lindblad
resonances plays a significant role in bringing these stars to us.  We will
see, however, that the mean radii of these stars are too large to be their
birth radii: even these stars have increased their angular momenta since
birth.

\end{itemize}

\renewcommand\arraystretch{1.5}
\begin{table*}%[htdp]
\caption{Median and interquartile values of the eccentricity distribution of the selected RAVE giants quality sample at three Galactocentric regions.}
\label{tab:interquartiles}
\begin{center}
\begin{tabular}{ccccccccc}
\hline \hline
{Metallicity range ($\dex$)} &$[-0.2,-0.1]$&$[-0.1,0.0]$&$[0.0,0.1]$&$[0.1,0.2]$&$[0.2,0.3]$ &  $[0.3,0.4]$ &  $[0.4,0.5]$ \\ \hline
 $6.5<R<7.5$& $ 0.18^{+0.28} _{-0.12}$ & $ 0.18^{+0.26} _{-0.11}$ & $ 0.19^{+0.28} _{-0.12}$ & $ 0.18^{+0.26} _{-0.11}$ & $ 0.19^{+0.28} _{-0.12}$ & $ 0.20^{+0.29} _{-0.12}$ & $ 0.18^{+0.24} _{-0.13}$ \\
$7.5<R<8.5$& $ 0.14^{+0.21} _{-0.09}$ & $ 0.14^{+0.21} _{-0.09}$ & $ 0.14^{+0.22} _{-0.09}$ & $ 0.13^{+0.20} _{-0.08}$ & $ 0.16^{+0.23} _{-0.10}$ & $ 0.18^{+0.25} _{-0.12}$ & $ 0.14^{+0.23} _{-0.09}$ \\
$8.5<R<9.5$ & $ 0.16^{+0.24} _{-0.10}$ & $ 0.17^{+0.24} _{-0.11}$ & $ 0.18^{+0.25} _{-0.11}$ & $ 0.16^{+0.27} _{-0.10}$ & $ 0.19^{+0.29} _{-0.13}$ & $ 0.16^{+0.25} _{-0.12}$ & $ 0.10^{+0.25} _{-0.07}$ \\ \hline
\end{tabular}
\end{center}
\end{table*}%

\renewcommand\arraystretch{1}
\begin{table*}%[htdp]
\caption{Ratio of circular orbit giant stars in the selected RAVE quality sample at three Galactocentric regions.}
\label{tab:radial_migrators}
\begin{center}
\begin{tabular}{ccccccccc}
\hline \hline
\multicolumn{2}{c}{Metallicity range ($\dex$)} &$[-0.2,-0.1]$&$[-0.1,0.0]$&$[0.0,0.1]$&$[0.1,0.2]$&$[0.2,0.3]$ &  $[0.3,0.4]$ &  $[0.4,0.5]$ \\ \hline
 &$6.5<R<7.5$& $0.38$ & $0.39$ & $0.36$ & $0.39$ & $0.35$ & $0.35$ & $0.31$ \\ %MEAN=0.36, sigma=0.03
${\rm N_{circ}/N_{total}}$ &$7.5<R<8.5$& $0.54$ & $0.55$ & $0.53$ & $0.57$ & $0.46$ & $0.40$ & $0.52$ \\ %\hline  %MEAN = 0.51, sigma=0.06
&$8.5<R<9.5$ & $0.47$ & $0.42$ & $0.41$ & $0.41$ & $0.33$ & $0.47$ & $0.60$ \\ \hline %MEAN= 0.44 , sigma=0.08
\end{tabular}
\end{center}
\end{table*}%

Based on the values presented in Table~\ref{tab:interquartiles} and illustrated in Fig.~\ref{fig:eccentricities}, 
 the central row of Table~\ref{tab:radial_migrators}
gives for
different metallicity bins the proportion of giant stars of a given
metallicity in the Solar neighbourhood that are on nearly circular orbits ($e
\lesssim 0.15$).  This ratio proves to be roughly constant, of the order of
0.51.  This is in agreement with \citet{Lee11}, who found that the thin
disc stars have eccentricities that are independent of metallicity. In
particular, we find that super-solar metallicity stars are not distributed
differently in either eccentricity and or mean orbital radius from the stars
of lower metallicity.

Taking into account {(i)} that super-solar metallicity stars are more metal-rich
than the local ISM, {(ii)} that they are not a young population and {(iii)}
that they are on nearly circular orbits, we can infer that these stars have
increased their angular momenta through resonant scattering at CR. 
{
A quantitative theoretical estimate of the probable birth radii of solar vicinity stars with $\meta > 0.2\dex$ is given by \citet{Minchev13,Minchev14b}. According to their simulations these SMR stars would originate from the $3-5\kpc$ galactocentric region and would be at most $6\Gyr$ old. From the observational point of view, we can say that the present abundance gradient in the ISM, according to the most precise estimates based on Cepheids \citep[e.g.][]{Genovali14}, is around $-0.06\dex\kpc^{-1}$. Although it is very unlikely that the gradient remained constant in the last $6-8\Gyr$, this would imply that $\meta = +0.4$ could be reached already at $2\kpc$ from the Galactic centre under the assumption of a linear gradient.

Figure~\ref{fig:ISM_gradients} represents an illustration of the possible birth radii of
any metallicity star, for two local metallicity
gradients in the ISM, considering the extreme case in which $\meta$
increases exponentially inwards with a scale length $R_{\rm M}$:
\begin{equation}\label{eq:metallicity_gradient}
\meta(R)=A\left(1-{\rm e}^{-(R-R_0)/R_{\rm M}}\right).
\end{equation}
The constant $A$ in this formula is the value to which $\meta$ tends at large
radii, and together with the value of  local metallicity gas gradient, sets the value of $R_{\rm M}$. Regardless of the adopted value of $A$, the shallowness of the local metallicity gradient in
the ISM always implies $R< 3.5\kpc$ for stars as metal-rich as $\meta=0.4$
}

\begin{figure}
\centering
\includegraphics[width=\linewidth, angle=0]{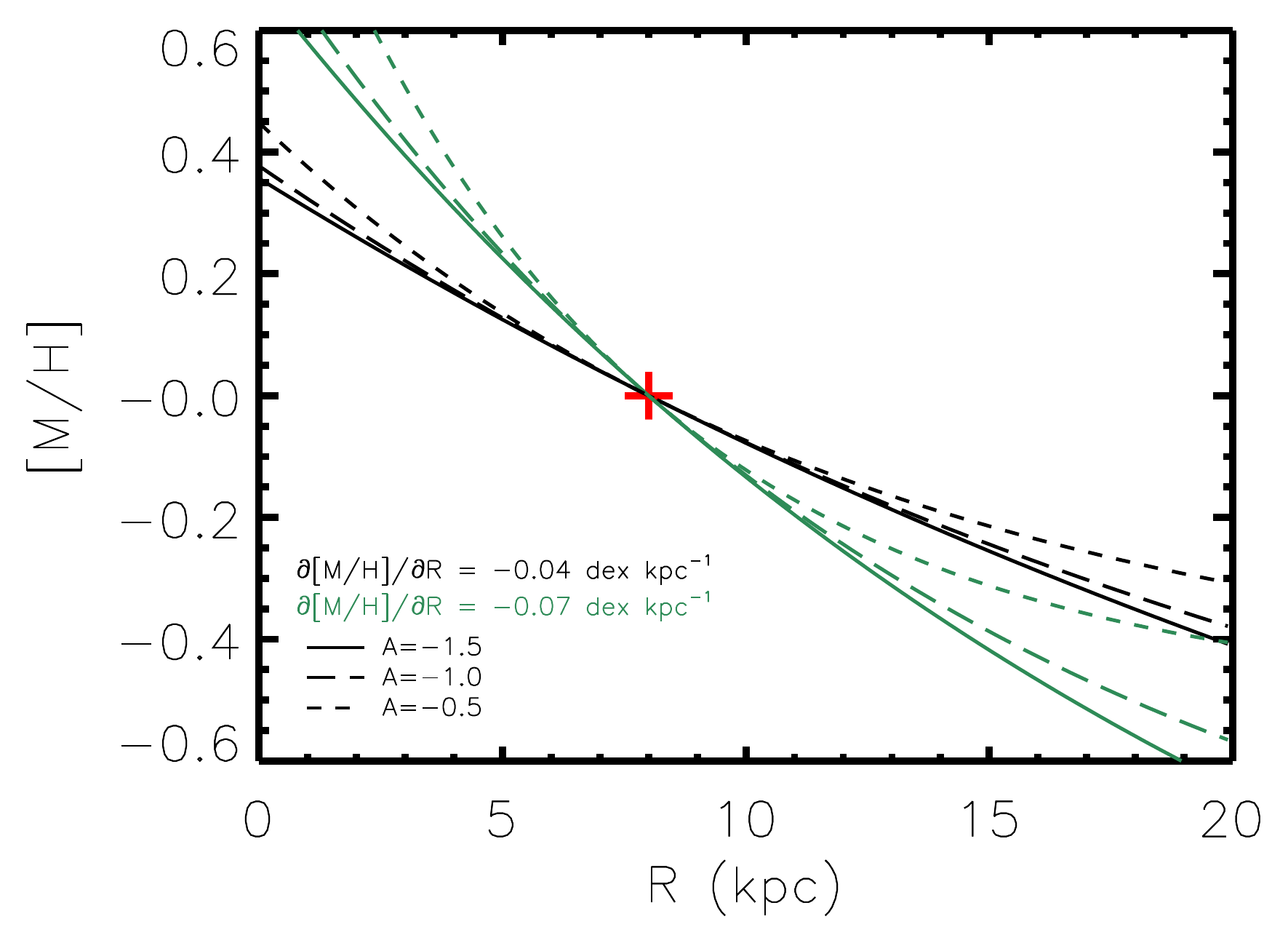}
 \caption{
 Example of an extreme steepening of the metallicity gradient of the ISM, following an exponential form (see Eq.~\ref{eq:metallicity_gradient}). Different metallicity values at high radii (factor $A$) and  locally measured ISM metallicity gradients are adopted to illustrate possible ranges of birth locations of the stars. The red '+' sign is at $(R_0,\meta)=(8\kpc,0)$. 
 }
\label{fig:ISM_gradients}
\end{figure}

%%%%%%%%%%%%%%%%%%%%%%%%%%%%%%%%%%%%%%%%%%%%%%
%%%%%%%%%%%%%%%%%%%%%%%%%%%%%%%%%%%%%%%%%%%%%%
The first and third rows of, Table~\ref{tab:radial_migrators} give ${\rm N_{circ}/N_{total}}$ for the cylindrical shell inside and outside
the solar cylinder, respectively. In these shells RAVE does not see stars that lie close to
the plane (Fig.~\ref{fig:metal_rich_map}), so only warmer populations are
sampled. Hence it is no surprise that in these shells we find smaller values
of ${\rm N_{circ}/N_{total}}$ than in the solar cylinder: in the inner
shell we have ${\rm N_{circ}/N_{total}}\approx 0.36$ while in the outer
shell we have ${\rm N_{circ}/N_{total}}\approx 0.44$.  However, the
measured ratios still show no statistically significant variation over the
entire investigated metallicity range (see
Figs.~\ref{fig:eccentricities},\ref{fig:mean_radii}).
%%%%%%%%%%%%%%%%%%%%%%%%%%%%%%%%%%%%%
 Below we discuss how the lack of dependence of ${\rm N_{circ}/N_{total}}$ on [Fe/H], even well
off the plane can put  constraints in the Galaxy's spiral history. 

The gravitational field of a spiral structure that has radial wavenumber $k$ varies with $z$ as $\exp(-k|z|)$
\citep[e.g.][\S6.1.5]{BT08}, so the capacity of a wave to force stars does
not extend further than $1/k$ from the plane. 
Among the recent literature, disc simulations of spirals with different radial wavenumbers resulted to different observational predictions on radial migration.  
For example, the disc simulations of \citet{Solway12} were seeded with spiral arms through the groove mechanism of
\citet{Sellwood91} and in these simulations the extent of radial
migration decreased only slowly with the amplitude of vertical excursions,
and was ``almost as large for thick-disc stars as for those in the thin disc''. 
On the other hand, in the simulations of \citet{Vera-Ciro14} the spirals were seeded by
point masses in the disc intended to represent giant molecular clouds, and it
was found, by contrast, that migrated stars were ``a heavily biased subset of
stars with preferentially low vertical velocity dispersions''. 
This finding reflects the short-wavelength, filamentary nature of the spiral structure
in the discs of \citet[][see their Fig.~2]{Vera-Ciro14}, which strongly confines
the gravitational field to the equatorial plane.

The lack of dependence of
${\rm N_{\rm circ}/N_{\rm total}}$  that we measure in our Galaxy implies that the responsible
spiral structure has longer radial scales than that discussed by \citet{Vera-Ciro14}. This is consistent with K-band photometry of nearby face-on spiral galaxies that shows that
the mass-bearing populations of these galaxies are organised into 
loosely-wound spirals  \citep[][Plate 1]{Rix95}.
A spiral structure, possibly groove driven like our results suggest, is also in agreement with \cite{Sellwood14b} that have argued that the groove mechanism
is fundamental to the development of large-amplitude spiral structure in
discs that contain only stars and are stable at the level of linear
perturbation theory.

%%%%%%%%%%%%%%%%%%%%%%%%%%%%%%%%%%%%%
\section{Discussion and Conclusions}
\label{sect:conclusions}

A recalibration of the metallicities of stars in the RAVE survey leads to a
significant increase in the number of stars with $\meta>0.2\dex$ and brings the
metallicity distribution of the entire RAVE sample into closer agreement with
that of the DR4 chemical pipeline. In fact the two distributions are now in
good agreement for $\meta>0.1\dex$.

Currently the inter-stellar medium (ISM) has metallicity $\meta<0$ near the
Sun, and its metallicity increases towards the Galactic centre, the local gas 
gradient being of the order of $\d\meta/\d R \simeq -0.06\dex\kpc^{-1}$
\citep[e.g.][]{Smartt97,Balser11,Genovali14}.  In the absence of large amounts of metal-poor gas being accreted,
the natural expectation for gas 
metallicity at a given radius is to be a monotonic increasing function of time,
so current metallicities give upper limits on metallicities at all previous
times \citep[e.g.][]{Chiappini09}. 
{  If these propositions are accepted and one grants that the metallicity
of a star reflects the metallicity of the ISM at the time and place of its
formation, it follows that super metal-rich (SMR) stars with
$\meta \sim 0.4\dex$ must have formed far inside $R_0$, probably in the 
region now occupied by the bar/bulge.

%%%%%%%%%%%%%%%
Indeed, the alternative scenario of these stars forming \emph{in situ}
nearer the Sun, from a turbulent ISM that has
reached $\meta \sim 0.4\dex$ at early times seems unlikely, because
turbulence extensive enough to mix gas from the inner galaxy to $R=6\kpc$ (where the stellar metallicity gradient measured for example by the APOGEE survey becomes flatter)
requires clouds to be on significantly non-circular orbits, and in such clouds few
stars would be born onto the near-circular orbits on which we observe many
SMR stars. Moreover, realistic
turbulent discs as seen in external galaxies at $z\sim1-2$
\citep[][]{Epinat12,Tacconi13} are too short lived \citep[e.g.][]{Genzel08,Bournaud09}
to allow enriched inner galactic ISM to reach outer galactic regions. 
%%%%%%%%%%%%%%%

We have shown that SMR stars are not on particularly eccentric orbits.  It
follows that they have materially increased their angular momenta
through resonant scattering by spiral arms \citep{Sellwood02}, and most no longer visit the
region of their birth. Angular momentum can be increased at either corotation
resonance (CR) or outer Lindblad resonance (OLR) and the key distinction
between these resonances is that at OLR eccentricity increases, while at CR
it does not. Since the SMR stars have experienced large angular momentum
increases without large increases in eccentricity, it follows that the
dominant process bringing them to us is scattering at CR, i.e., churning.

}

We have shown that the vertical distribution of the higher-metallicity stars
is not unusual. In fact, as far as we can determine, the spatial distribution
of these stars is independent of metallicity.  The natural interpretation of
this finding is that the probability for radial migration is insensitive to
the extent of a star's excursions perpendicular to the plane. This
interpretation is consistent with the dynamical study of \cite{Solway12}, who
showed an example of a disc seeded with spiral structure through the groove
mechanism of \cite{Sellwood91} where migration probability is almost as
large for thick-disc stars as for those in the thin disc. 
Our interpretation
of the data implies that the radial wavelength of spiral structure is no
smaller than the thick disc's scale height, which is not the case in the
experiments of \cite{Vera-Ciro14} who used a simulation with  multi-armed recurrent spirals and showed that churning is far more efficient for stars with low vertical velocity dispersion.

Finally, we note that rather than radial migration, the SMR stars could be
evidence of inhomogeneous chemical evolution within the Galactic disc on
account of the Galactic fountain \citep[e.g.:][and references
therein,]{Marasco13} bringing metal-rich ejecta from an
interior region of the disc to somewhere in the solar annulus and there
giving rise to metal-rich star formation \citep{Spitoni09}. However, this
scenario seems implausible because (i) the ratio of stars on nearly circular
and eccentric orbits is independent of metallicity, and (ii) high-velocity
clouds, which may be associated with the Galactic fountain, have low
metallicities \citep{Wakker01}. However, only a few hundred high-resolution
spectra of SMR stars should be enough to identify their chemical abundance
patterns and test this possibility.  Large and high-resolution surveys such
as Gaia-ESO \citep{Gilmore12} or APOGEE \citep{Allende-Prieto08} could,
perhaps, already provide some answers to the above questions.

%%%%%%%%%%%%%%%%%%%%%%%%%%%%%%%%%%%%%%%%%
\section*{Acknowledgments}
The anonymous referee is greatly thanked for the useful comments and suggestions that helped improving the quality of this paper. 
 Funding for RAVE has been provided by: the Australian
Astronomical Observatory; the Leibniz-Institut f\"ur Astrophysik Potsdam (AIP); the Australian National University;
the Australian Research Council; the French National Research Agency; the German Research Foundation (SPP 1177
and SFB 881); the European Research Council (ERC-StG
240271 Galactica); the Instituto Nazionale di Astrofisica at
Padova; The Johns Hopkins University; the National Science Foundation of the USA (AST-0908326); the W. M.~Keck foundation; the Macquarie University; the Netherlands Research School for Astronomy; the Natural Sciences
and Engineering Research Council of Canada; the Slovenian Research Agency; the Swiss National Science Foundation; the Science \& Technology Facilities Council of the UK;
Opticon; Strasbourg Observatory; and the Universities of
Groningen, Heidelberg and Sydney. 
RFGW acknowledges support of NSF Grant OIA-1124403 and thanks the Aspen Center for Physics and NSF Grant \#1066293 for hospitality during the  writing of this paper. The research leading to these results has received funding from the European Research
Council under the European Union's Seventh Framework Programme (FP7/2007-2013)/ERC
grant agreement no.\ 321067.
The RAVE web site is
at \url{http://www.rave-survey.org}.

\bibliographystyle{mn2e}
\def\aj{AJ}\def\apj{ApJ}\def\apjl{ApJL}\def\araa{ARA\&A}\def\apss{Ap\&SS}
\def\mnras{MNRAS}\def\aap{A\&A}\def\nat{Nature}
\def\nar{New Astron. Rev.}

\bibliography{MR_stars}

\label{lastpage}

\end{document}